# Non-Equilibrium Photodissociation Regions:
# Ionization-Dissociation Fronts

Frank Bertoldi

Max-Planck-Institut für Extraterrestrische Physik, D-85740 Garching, Germany

fkb@mpe-garching.mpg.de

and

B.T.Draine

Princeton University Observatory, Peyton Hall, Princeton, NJ 08544, USA

draine@astro.princeton.edu

## ABSTRACT

We discuss the theory of coupled ionization–dissociation fronts produced when molecular clouds are exposed to $\lambda < 1110$Å radiation from hot stars. A steady, composite structure is developed, which generally includes an ionized outflow away from the cloud, an ionization front, a layer of photodissociated gas, a photodissociation front, and a shock wave preceding the photodissociation front. We show that the properties of the structure are determined by two dimensionless parameters, $\psi$ and $\delta$, and by the Alfvén speed in the preshock gas. For a broad range of parameters of interest, the ionization front and the hydrogen photodissociation front do not separate, the $H_2$ photodissociation and photoionization take place together, and a classical hydrogen "photodissociation region" (PDR) does not exist. We also show that even when a distinct photodissociation region exists, in many cases the dissociation front propagates too rapidly for the usual stationary models of PDRs to be applicable. We discuss several famous PDRs, e.g., in M17 and Orion and conclude that they cannot be described by equilibrium PDR models.

*Subject headings:* ISM: molecules – molecular processes – galaxies: ISM – nebulae: H II regions, Orion Nebula – shock waves

## 1. Introduction

A photodissociation region (PDR) is the transition layer at the surface of a molecular cloud between the dense, cold molecular gas and the tenuous, warm, ionized gas; it is the region where the far-ultraviolet (FUV) photons in the range $912 - 1110$ Å are either absorbed by $H_2$ (which in $\sim 15\%$ of the cases leads to dissociation) or by dust. Over the last decade observations have shown that a significant fraction of the energy output from massive stars is absorbed and reprocessed in PDRs, and they have received increased theoretical attention [see the review by Hollenbach & Tielens (1994), and references therein]. The quantitative comparison of PDR models with observations can yield important information on the chemical and topological structure of molecular clouds and the effects of massive star formation on them.

The sharp transition where molecular hydrogen turns mostly atomic we call the photodissociation front (DF); it will propagate into the molecular cloud when the incident FUV flux increases or the self-shielding neutral gas and dust layer (i.e., the PDR) evaporates off the surface of the cloud. This occurs when the atomic gas layer is eroded by a propagating photoionization front (IF) at which the neutral gas is ionized by the Lyman continuum (Lyc: $\lambda < 912$Å) radiation and, because of its sudden heating to $10^4$ K, accelerates away from the cloud surface. Note that the most prominent PDRs, i.e. those receiving a strong UV flux, are usually also subject to erosion by Lyc radiation, and that therefore the structure and dynamics of such PDRs cannot be considered without including the effects of the ionizing radiation.

Semantic differences exist in the definition of what constitutes a PDR. The acronym is also used for "photon-dominated region," which typically denotes a surface layer that reaches deeper into a molecular cloud, and in which, despite its exponential attenuation, the FUV radiation dominates the chemical and thermal structure of the gas. In fact, most of the cold ISM is dominated by FUV ionization (e.g. McKee 1989), and only a small fraction of molecular gas is sufficiently shielded that cosmic rays are the main source of ionization.

To distinguish the cold, low-ionization gas from the infrared–bright surface layers where most of the UV radiation is absorbed, for us "PDR" shall denote only the neutral H layer surrounding a cold molecular cloud where $H_2$ has recently been dissociated by FUV photons. The PDR is bounded by a H ionization and a $H_2$ dissociation front. The photodissociation of molecules such as CO can occur upstream of the $H_2$ dissociation front, and some of the characteristic emission associated with PDRs (from e.g. O, C, or $C^+$) may arise there.

Most theoretical PDR models so far assumed the gas to be at rest with respect to the DF, and most (e.g. Tielens & Hollenbach 1985; Burton et al. 1990; etc.) assumed a steady-state balance between $H_2$ destruction (mainly by UV pumping) and $H_2$ formation (mainly by grain catalysis).

Time-dependent PDRs have previously been discussed by Hill & Hollenbach (1978) and Roger & Dewdney (1992), who studied the interaction of a spherical ionization-shock front with the DF in an expanding H II region. London (1978) and Goldshmidt & Sternberg (1995) follow the propagation of a DF into a planar cloud when the radiation source first turns on. They find that advancing DFs produce larger column densities of vibrationally excited $H_2$ than stationary ones because grains initially absorb a smaller fraction of the UV photons. Wagenblast & Hartquist (1988) also studied the changing $H_2$ level population when the DF first penetrates a translucent cloud.

Giant molecular clouds are known to be clumpy, with most of the molecular mass confined to clouds (or clumps) that fill only a small fraction of the volume of a GMC (e.g. Bertoldi & McKee 1992; Williams & Blitz 1995). When an O star creates an H II region within a clumpy GMC, most of the ionization and dissociation therefore occurs on the surfaces of the small, dense molecular clouds. None of the previous works considered the interaction of an ionization-shock front with a dissociation front while both fronts penetrate such a spheroidal molecular cloud. In this paper we show that the flow of the gas through the DF, PDR, IF, and off the cloud surface can have serious effects on the PDR structure not only when a cloud is first exposed to the radiation, but throughout the lifetime of the massive star or cloud. The effects of advection on the structure and emission characteristics of PDRs must therefore receive special attention in future PDR models.

In section 2 we broadly specify for which astrophysical objects we expect the IF to have little effect on the structure of the PDR. In section 3 we introduce the two main parameters that define the impact of the ionizing radiation on a cloud: the photoevaporation parameter, $\psi$ is a measure for the Lyc opacity of a steady photoevaporation flow off the surface of the cloud, and the Strømgren parameter, $\delta$, is a measure for the dynamical impact of the photoevaporation on the cloud. In section 4 we compute the propagation speeds of the ionization and $H_2$ dissociation fronts and determine in which cases both fronts form a merged structure, so that an extended neutral HI layer can not develop. We find that this condition mainly depends on the photoevaporation parameter of the cloud and the ratio of the Lyc and FUV emission of the star. Section 5 compares the flow time through a propagating DF with the timescale within which equilibrium $H_2$ abundances are established. We can thereby show in which cases transient photochemistry has serious effects on the structure of the PDR. In section 6 we discuss the implications of our findings for several well-known objects, including the PDR behind the Orion Trapezium stars, and section 7 summarizes our findings and discusses their implications.

## 2. Cases Unaffected by Ionizing Radiation

The structure of PDRs is most affected by ionizing radiation where the latter is strong compared to the FUV. We focus our discussion therefore on molecular clouds that are irradiated by one or more nearby O stars. We thereby exclude objects such as reflection nebulae, H I regions dissociated by the diffuse FUV background radiation, and the edges of certain H II regions.

### 2.1. Reflection Nebulae

B stars have very weak Lyc emission but can form prominent shells of warm, dissociated H I around their small H II regions; NGC 2023 is a famous example of such a "reflection nebula" (e.g. Gatley et al. 1986; Draine & Bertoldi 1996). The dynamics of such PDRs are not strongly affected by ionizing radiation.

### 2.2. Clumps in Molecular Clouds

Far infrared observations of e.g. $C^+$ (Stutzki et al. 1988) show that the diffuse FUV background radiation can apparently propagate deep into clumpy molecular clouds, dissociate the tenuous interclump gas and irradiate the surfaces of dense molecular clumps. The PDRs around such clumps are likely in pressure equilibrium with the interclump gas and are not affected by ionizing radiation. However, if the FUV in such a clumpy cloud is enhanced by nearby massive stars (as in the Rosette Molecular Cloud: Cox et al. 1990) and is increasing due to the approach of an ionization front, time-dependent effects may become important; the PDRs may begin to propagate into the clumps in response to the increasing FUV flux.

### 2.3. Edges of Spherical H II Regions

At the edge of an H II region that expands into a homogeneous molecular cloud (i.e., an ideal Strømgren sphere), the ionizing radiation is weak and the IF propagates only slowly into the PDR that forms the inner boundary to the dense neutral shell piled up during the expansion. If photoevaporating clumps inside the H II region contribute significantly to its mass input (McKee et al. 1984), the edge of the H II region can even become a recombination front (Bertoldi 1989a). In either case the ionizing radiation does not have a strong effect on the structure of the PDR, and quasi-equilibrium models (Hill & Hollenbach 1978; Roger & Dewdney 1992) are appropriate to describe the evolution of such PDRs, except during the initial phases of their expansion when the IF is R-type.

## 3. Photoevaporation Flow Parameters: $\psi$ and $\delta$

We now consider PDRs on the surfaces of molecular clouds which are illuminated by one or more O stars. We assume that the surface is convex, i.e., curved away from the star, so that the newly ionized gas can freely expand into the H II region (Fig.1). Let $r_c$ be the radius of curvature of the surface, and $R$ be the distance from the surface to the star. The structure of such a photoevaporation flow, the ionization front, and the compressive effect on the cloud were discussed in detail by Bertoldi (1989b) and Bertoldi & McKee (1990) for the case $R \gg r_c$. They showed that the photoevaporation of a spheroidal cloud is essentially characterized by two parameters. The dimensionless "photoevaporation parameter" $\psi$ is a measure for the opacity of the evaporation flow to Lyc photons:

$$\psi \equiv \frac{\alpha_B \, S_{ly} \, r_c}{4\pi R^2 c_2^2} = 5.15 \times 10^4 \, \frac{S_{49} r_{pc}}{R_{pc}^2} \, , \tag{1}$$

where $\alpha_B \simeq 2.59 \times 10^{-13} \mathrm{cm^3 s^{-1}}$ is the case B hydrogen recombination coefficient at $10^4 \, \mathrm{K}$, $c_2 \simeq 11.4 \, \mathrm{km \, s^{-1}}$ is the typical isothermal sound speed in the ionized gas, $r_c = r_{pc} \mathrm{pc}$, $R = R_{pc} \mathrm{pc}$, and $S_{ly} = 10^{49} S_{49} \, \mathrm{s^{-1}}$ is the emission rate of Lyc photons by the star. Typical photoevaporation flows of interest have $\psi \gg 1$.

Here we consider the more general case, including the limit of a planar cloud, $R \ll r_c$. Let $r$ be the distance from the ionizing source. Along the symmetry axis, the ionized gas streaming off the cloud surface has a density profile approximated by

$$n(x) \approx n_2 (x/r_c)^{-2.5} \, , \tag{2}$$

where $x = R - r + r_c$ is the distance from the center of curvature, and $n_2 = F_{ly}/v_2$, where $F_{ly}$ is the ionizing flux arriving at the cloud surface, and $v_2 \approx 1.6 c_2$ is the characteristic speed of the evaporation flow. The exponent 2.5 and coefficient 1.6 were chosen to fit the high–$\psi$ density profiles computed by Bertoldi (1989b).

In ionization equilibrium, the radiation transfer equation along a radial line from the star is

$$\frac{1}{r^2} \frac{d}{dr} \left[ r^2 F(r) \right] \; = \; -\alpha_B n(r)^2 \; - \; n(r) \sigma_{ly} \, F(r), \tag{3}$$

where $F(r)$ is the Lyc flux at some distance $r$ from the star, and $\sigma_{ly}$ is the dust <u>absorption</u> cross section per H nucleus for the ionizing photons. We adopt the "on-the-spot" approximation, assuming that recombination to the ground state and the resulting Lyman continuum photons can together be omitted. Similarly, scattering of Lyman continuum photons by dust is neglected, as these scattered photons remain available for ionizing hydrogen.

### 3.1. Dust Properties

Interstellar dust properties vary from region to region [see the review by Mathis (1990)]. The extinction longward of $I = 9000$Å appears to be "universal" (Cardelli, Clayton & Mathis 1989), with $A_I/N_H \approx 2.60 \times 10^{-22}\,\mathrm{cm}^2$ (Draine 1989). For $\lambda < 9000$Å, the extinction law varies, depending on the value of $R_V \equiv A_V/E(B-V)$. Diffuse cloud dust is characterized by $R_V \approx 3.1$, and the extinction cross section per H nucleus is estimated to be $\sim 2.6 \times 10^{-21}\,\mathrm{cm}^2$ at $1000$Å (Cardelli, Clayton & Mathis 1989). For dense clouds, however, the dust is characterized by larger values of $R_V \approx 4 - 6$ (Mathis 1990), and reduced FUV extinction. For the numerical examples in most of this paper we will consider dust characterized by an average dense cloud $R_V \approx 5$, close to the high value of 5.5 that is found in the Orion (M42) nebula (§6.3).

Grain models predict that the absorption will rise to a maximum near $\sim 750$Å, beyond which it will decline (Laor & Draine 1993). Since we are mostly concerned with hot stars in dense H II regions, the ionizing photons near the ionization front should have average photon energies well above $13.6$eV. Thus the effective extinction cross section for these photons is smaller than the maximum value at $800$Å. In lieu of reliable extinction models, we will assume that the average extinction cross section is close to that extrapolated to the Lyman edge, $\sim 1.22 \times 10^{-21}\,\mathrm{cm}^2/$H; with an estimated albedo of 0.5, the Lyc absorption cross section is then $\sigma_{ly} \approx 0.61 \times 10^{-21}\mathrm{cm}^2$.

### 3.2. Lyc Transfer Solution

The general solution to equation (3) is

$$r^2 F(r) = e^{-\tau(r)} \left[ \frac{S_{ly}}{4\pi} - \int_0^r dr'\; r'^2 \; \alpha_B n^2 e^{\tau(r')} \right] \;, \tag{4}$$

where

$$\tau(r) \equiv \sigma_{ly} \int_0^r n(r')dr' \;. \tag{5}$$

With our assumed density profile $n(r)$, and defining $\tau_{ly} \equiv \tau(R)$, $F_{ly} \equiv F(R)$, equation (4) becomes

$$\frac{S_{ly}}{4\pi R^2} e^{-\tau_{ly}} - F_{ly} = \frac{\alpha_B n_2^2 r_c}{4} \mathcal{I}(\tau_{ly}, R/r_c) ,  \qquad (6)$$

where

$$\mathcal{I}(\tau_{ly}, z \equiv R/r_c) \equiv 4 \int_1^{1+z} x^{-5} \left(1 - \frac{x-1}{z}\right)^2 \exp\left[-\tau_{ly}\frac{1 - x^{-1.5}}{1 - (1+z)^{-1.5}}\right] dx \qquad (7)$$

$$\approx (1 + \tau_{ly}/3)^{-1}(1 + 0.72/z)^{-1} . \qquad (8)$$

The approximation (8) has a maximum error of 8% near $z = 1$ and $\tau_{ly} = 3$ for $z \in [.01, 10^4]$ and $\tau_{ly} \in [0, 10]$.

A fraction of the incident Lyc photons is absorbed by dust and by the recombining H and He ions in the evaporation flow, and the rate of arrival of photons at the ionization front is reduced to $F_{ly} \equiv S_{ly}/4\pi R^2 q$, where the absorption factor, $q$, is given by

$$q\,(qe^{-\tau_{ly}} - 1) = 0.1\,\psi\,\mathcal{I}(\tau_{ly}, R/r_c) ,  \qquad (9)$$

which derives from multiplying equation (6), by $S_{ly}/[4\pi R^2 F_{ly}^2]$. Integrating (5) with (2) yields

$$\tau_{ly} = \frac{\sigma_{ly} c_2}{1.5 \times 1.6\,\alpha_B}\frac{\psi}{q}\left[1 - \frac{1}{(1+R/r_c)^{1.5}}\right]$$

$$= \frac{\psi}{545\,q}\left(\frac{\sigma_{ly}}{10^{-21}\,\mathrm{cm}^2}\right)\left[1 - \frac{1}{(1+R/r_c)^{1.5}}\right] , \qquad (10)$$

which is the dust absorption optical depth for Lyc photons through the freely expanding evaporation flow between the star and the ionization front. We may treat $R/r_c$ and $\tau_{ly}$ as the two parameters defining the problem, and find $q(\tau_{ly}, R/r_c)$ and $\psi(\tau_{ly}, R/r_c)$ from equations (9) and (10). With $\sigma_{ly} \equiv 10^{-21}\sigma_{-21}$ cm$^2$ we obtain

$$\psi = \frac{2.99 \times 10^4}{\sigma_{-21}^2}\frac{\tau_{ly}^2\,e^{\tau_{ly}}}{1 + \tau_{ly}/3} \times \begin{cases} (1 + 0.018\sigma_{-21}/\tau_{ly}) & \text{for } R \gg r_c, \\ 0.61\,(r_c/R) & \text{for } R \ll r_c, \end{cases} \qquad (11)$$

$$q = \left[1 + \frac{54.5\,\mathcal{F}(R/r_c)}{\sigma_{-21}}\frac{\tau_{ly}}{1 + \tau_{ly}/3}\right]\,e^{\tau_{ly}} , \qquad (12)$$

where

$$\mathcal{F}(R/r_c) \equiv (1 + 0.72 r_c/R)^{-1}\left[1 - (1+R/r_c)^{-1.5}\right]^{-1} = \begin{cases} 1 & \text{for } R \gg r_c, \\ 0.93 & \text{for } R \ll r_c. \end{cases} \qquad (13)$$

The photoevaporation flow is optically thick to Lyc radiation for $q \gg 1$, i.e. $\psi \gg 10$. Known evaporation flows have values of $\psi$ up to $\sim 10^5$ and $\tau_{ly} \approx 1$ for, e.g., the "partially ionized globules" (PIGs) in the Trapezium cluster.

### 3.3. Strømgren Number and Shock Speed

Our second dimensionless parameter, the "Strømgren number"

$$\delta \equiv \frac{S_{ly}/4\pi R^2}{2\alpha_B \, n_0^2 \, r_c} \;=\; 5.23 \times 10^{-4} \, \frac{S_{49}}{R_{pc}^2 r_{pc}} \left( \frac{10^4 \, \text{cm}^{-3}}{n_0} \right)^2 \;, \tag{14}$$

is the ratio of the width $(S_{ly}/4\pi R^2)/(\alpha_B n_0^2)$ of an ionized Strømgren layer of density $n_0$ (in which recombinations balance ionizations), and the cloud size, $2r_c$. A cloud of initial density $n_0$ and $\delta > 1$ (and $\delta\psi > 1$) that is first exposed to the ionizing radiation gets completely ionized by a rapid R-type ionization front that never stalls to develop a shock. When $\delta$ is small however, only a thin surface layer is ionized before a shock forms and the shielding photoevaporation flow is established. For large $\psi$, the velocity of the shock is derived from the momentum jump condition. If we assume a strong shock, $v_s \gg v_{A0}$, where $v_{A0}$ is the preshock Alfvén speed, then $\rho_0 v_s^2 \simeq \rho_2(v_2^2 + c_2^2)$; using the identity $S_{ly}/4\pi R^2 n_0 c_2 = (2\psi\delta)^{1/2}$ and equation (9) with $q \gg 1$, we obtain

$$v_s \simeq c_2 \, (99 \, \delta)^{1/4} e^{-\tau_{ly}/4} \, (1 + \tau_{ly}/3)^{1/4} (1 + 0.72 r_c/R)^{1/4} \;. \tag{15}$$

However, $v_s$ cannot exceed $c_2$, since otherwise the evaporating gas could not diverge as it leaves the IF and would accumulate between the cloud and the star, leading to increasing absorption and the deceleration of the shock (Bertoldi 1989b).

If the shock compression is limited by the cloud's magnetic field, as is most likely the case for typical interstellar conditions, then the shock-compressed gas has a density

$$n_1 \approx \sqrt{2} \, \frac{v_s}{v_{A0}} \, n_0 \;, \tag{16}$$

where $v_{A0} = v_{A0,5} \text{km s}^{-1}$ is the Alfvén speed in the unperturbed, upstream gas; the Alfvén speed in the compressed gas is $v_{A1} \simeq 2^{1/4}(v_s v_{A0})^{1/2}$. The clumps in molecular clouds show linewidths that suggest $v_{A0,5} \approx 1 - 3$ (e.g. Bertoldi & McKee 1992). The IF propagates into the shocked gas layer at the velocity

$$v_{IF} = \frac{S_{ly}}{4\pi R^2 q n_1} \approx \; v_{A0} \, \delta^{1/4} \, e^{-\tau_{ly}/4} \, (1 + \tau_{ly}/3)^{1/4} (1 + 0.72 r_c/R)^{1/4} \;, \tag{17}$$

following a derivation similar to that of (15).

The pressure of the ionized gas that freely streams off the cloud surface would not suffice to drive a shock into the cloud if $\delta \lesssim 10^{-6} v_{A0,5}^4$. However, such a situation would not occur in a cloud where the clumped, dense molecular gas was in pressure equilibrium with the tenuous interclump gas before an H II region developed. This can be seen from the following consideration. When an H II region develops that engulfs a denser molecular clump, the pressure of the more tenuous interclump gas certainly rises as it becomes ionized. If the clump was initially in pressure equilibrium with the interclump gas, its enhanced pressure would now lead to compression of the clump. However, the pressure of a freely expanding evaporation flow would, if $\delta \lesssim 10^{-6} v_{A0,5}^4$, be lower than the initial cloud pressure, and thus also lower than the pressure of the ionized interclump gas. We conclude that a steady, freely expanding evaporation flow could not form. Furthermore, it is easy to show that under such conditions the interclump gas is sufficiently dense to absorb all Lyc photons before they even reached the clump's surface, unless the star was located very close to the clump surface. Thus for very small values of $\delta$, the star's Strømgren sphere in the interclump gas would be smaller than the clump.

## 4. Merged Fronts: Ionization Erodes Dissociation Region

Consider a molecular cloud of H nucleon density $n_1$ and effective radius of curvature $r_c$ that is suddenly exposed to a FUV (912-1110Å) flux $F_{uv} = (S_{uv}/4\pi R^2)e^{-\tau_{uv}} = \chi F_H$, where $F_H = 1.21 \times 10^7 \text{cm}^{-2}\text{s}^{-1}$ is the flux for the diffuse UV spectrum of Habing (1968) in this wavelength interval, and $\tau_{uv} = (\sigma_{uv}/\sigma_{ly})\tau_{ly}$ is the UV optical depth due to dust in the ionized gas between the molecular cloud surface and the star; $\sigma_{uv}$ is the effective dust <u>attenuation</u> cross section for FUV photons. For $R_V = 5$ the dust <u>extinction</u> cross section at 1000Å is $\sigma_{ext} \approx 1.02 \times 10^{-21}\text{cm}^2$. We assume a dust albedo $\sim 0.5$ in the UV, and estimate that the effect of scattering is such that the effective attenuation cross section is $\sigma_{uv} \simeq \sigma_{abs} + 0.5\sigma_{sca} \simeq 0.75\sigma_{ext} \approx 0.76 \times 10^{-21}\text{cm}^2$, and $s \equiv \sigma_{uv}/\sigma_{ly} \simeq 0.76/0.61 = 1.25$. Note that $s$ varies slightly with $R_V$: $s \approx 1.20$, $1.25$, and $1.30$ at $R_V = 3.1$, $5$, and $6$, respectively. The corresponding values for $(\sigma_{uv}/10^{-21}\text{cm}^2)$ are $1.64$, $0.61$, and $0.38$.

When the dissociating radiation first strikes the cloud surface, a dissociation front propagates into the cloud at the velocity

$$v_{DF} = 2\,\epsilon\,f_d\,F_{uv}\,/\,n_1 = 18.2\,\frac{\epsilon}{0.5}\,\frac{f_d}{0.15}\,\frac{\chi}{n_1}\,\text{km s}^{-1}, \tag{18}$$

where $\epsilon \simeq 0.5$ is the fraction of the incident FUV photons which are absorbed by $H_2$, accounting for the fact that about half the photons are far from strong pumping lines and are thus absorbed by dust (Draine & Bertoldi 1996); $f_d \simeq 0.15$ is the fraction of $H_2$ pumpings that lead to dissociation. As the DF propagates into the cloud, ever more UV photons are absorbed in the dissociated layer between the ionization and dissociation fronts (i.e., the region we term the PDR) by dust and the reformed $H_2$. As a consequence, the DF slows down, and, if there were no erosion of the dissociated atomic layer by ionizing radiation, would eventually stall when the PDR has attained an equilibrium between $H_2$ destruction and formation. The initial propagation of a DF into a static cloud has recently been studied by Goldschmidt & Sternberg (1995).

Since the cloud is also exposed to ionizing radiation, an IF penetrates the cloud surface at the velocity $v_{IF} = S_{ly}/4\pi R^2 qn_1$. Initially $q = 1$ since no recombining dense gas shields the cloud surface, but after a short time $\sim \min(r_c, R)/2c_2 \simeq 4 \times 10^4 \; \min(r_{pc}, R_{pc})$ yr a dense evaporation flow is established and $q$ is implicitly given by equations (11) and (12).

Whenever ionizing and dissociating radiation strikes the surface of a molecular cloud, both the IF and DF begin to propagate into the cloud. *When $v_{IF} > v_{DF}$, the DF cannot separate from the IF and the ionization and dissociation fronts form a merged structure; a PDR will not be able to form.* With the absorption factor $q$ from equation (12), the condition $v_{IF} > v_{DF}$ becomes

$$\frac{0.5}{\epsilon} \frac{0.15}{f_d} \frac{S_{ly}}{S_{uv}} > 0.15 \left[1 + \frac{54.5 \; \mathcal{F}(R/r_c)}{\sigma_{-21}} \frac{\tau_{ly}}{1 + \tau_{ly}/3}\right] e^{-(s-1)\tau_{ly}}. \tag{19}$$

For $\sigma_{-21} = 0.61$ and $s = 1.25$, the RHS of equation (19) (Fig.2) attains a maximum value 9.9 (9.2) for $R/r_c \gg 1 (\ll 1)$ at $\tau_{ly} = 2.26$. For the assumed values of $f_d$ and $\epsilon$, the IF will therefore *always* be faster than the DF when $(S_{ly}/S_{uv}) > 9.9(9.2)$, but such high values of $S_{ly}/S_{uv}$ are rarely the case for a stellar radiation field. When the LHS is smaller than 9.9 (9.2), there will be a regime between a lower (provided the LHS exceeds 0.15) and upper $\tau_{ly}$ between which the inequality (19) is satisfied so that the DF is faster than the IF and a PDR can form. Larger values of $\sigma_{-21}$ or $s$ would enlarge the parameter space where the fronts merge (cf. Fig.2).

The low $\tau_{ly}$ regime of overlap arises because with insignificant absorption in the evaporation flow, the IF will propagate faster than the DF when the flux of incident ionizing photons is higher than twice the flux of UV photons that lead to dissociation (estimated as $\sim 7.5\%$ of all incident UV photons). With increasing $\tau_{ly}$, the Lyc radiation is absorbed by both dust and recombined H – the UV radiation however is only absorbed by dust, resulting in an increase in the velocity of the DF relative to the IF. At high $\tau_{ly}$ and $s > 1$, the difference between the effective $\sigma_{uv}$ and $\sigma_{ly}$ results in $v_{IF}/v_{DF}$ increasing exponentially with $\tau_{ly}$. The larger the value of $s = \sigma_{uv}/\sigma_{ly}$, the lower the value of $\tau_{ly}$ at which the IF is again faster than the DF (cf. Fig.2).

We computed the main sequence lifetime–averaged Lyc and FUV luminosities of O stars by combining the stellar evolution calculations of Schaller et al. (1991) with the non-LTE stellar atmosphere models of Kunze (1994). For stars of masses between 20 (O9) and $60 M_\odot$ (O5), we find

$$\frac{S_{ly}}{S_{uv}} \simeq 1.0 \; S_{49}^{0.5} \qquad \text{for } .085 \lesssim S_{49} \lesssim 3 \; , \qquad (20)$$

and this ratio levels off for higher masses to reach 2.2 for a star with initially $120 M_\odot$ (Kunze & Bertoldi, in preparation). For a molecular clump exposed to the ionizing and dissociating radiation of a single O5 star (for which $S_{49} \simeq 3$), we expect the IF and DF to overlap for values of $\tau_{ly}$ below 0.24 or above 4.2, whereas for a $20 M_\odot$ O9 star with $S_{49} \simeq 0.085$ overlap occurs only when $\tau_{ly} < .018$ or $\tau_{ly} > 8.3$. In Fig. 3a the vertical lines delineate the areas in parameter space to the very left and very right where overlap would occur. We are however not aware of any observed evaporation flows that are thick enough as to fall into the high $\tau_{ly}$ regime of overlap.

Note that for massive stars the surface temperature, and thereby the ratio of Lyc to UV emission can vary significantly during the evolution of the star. Young stars tend to have higher effective surface temperatures and lower surface metallicities in their photospheres, resulting in higher Lyc to UV emission ratios that gradually decline during the main sequence evolution. WR stars may also have higher than MS–average Lyc/UV emission ratios.

If the radiation incident on the clump is due to a cluster of OB stars, the most massive star may dominate the Lyc irradiation, but the less massive stars may contribute significantly to the UV flux. An integration of the UV and Lyc emission over a stellar mass distribution similar to the field star IMF shows that the ratio of the integrated Lyc to UV emission still exceeds unity if the most massive star is earlier than O5. However, the relative strength of the UV and Lyc radiation on the cloud surface depends on the distances to the individual stars.

## 5. Non-Equilibrium PDR: Transient Photochemistry

A PDR is the neutral gas layer between the IF and DF where the abundance of $H_2$ is low due to photodissociation by FUV photons. Stationary models of PDRs (which assume $v_{DF} = 0$) postulate a steady-state balance between $H_2$ destruction (mainly by UV-pumping) and $H_2$ formation (mainly by grain catalysis) at each point. Grain catalysis of $H_2$ is slow, with a timescale $t_{gr} = (2R_{gr} n_H)^{-1} \simeq 5 \times 10^8 \; \text{yr}/(n_H / \text{cm}^{-3})$ (for $R_{gr} = 3 \times 10^{-17} \, \text{cm}^3 \, \text{s}^{-1}$; Jura [1975]). Hence the local rate per volume of UV pumping of $H_2$ must everywhere be low. In stationary models of thick PDRs ($\chi/n \gg 10^{-2} \, \text{cm}^{-3}$), the FUV is attenuated in a layer of dusty gas with very low $H_2$ abundance before reaching the molecular region.

In an advancing PDR ($v_{DF} > 0$), however, steady-state conditions do not apply, and there will be regions in which the local $H_2$ pumping rate per volume is increased (Goldshmidt & Sternberg 1995). This increase is likely to heat the gas beyond its equilibrium temperature (London 1978) and can alter the chemical structure in the PDR.

Let $L$ be the width of the layer in which the transition from atomic to molecular H occurs, i.e., the width of the DF. The importance of "transient photochemistry" can be assessed by comparing the flow time through the DF, $L/v_{DF}$, with the timescale of $H_2$ formation on grains, $t_{gr}$. For this comparison to be meaningful we need to assume that the PDR exists, i.e., that the DF and IF are not merged and inequality (19) is not met. Then the IF and DF both propagate into the molecular cloud at the same speed, separated by a layer of mostly neutral and warm gas, the PDR. When the cloud is first exposed to the strong FUV and Lyc radiation, the DF may propagate more rapidly than the IF, but as the FUV flux arriving at the DF is attenuated by the dust and self-shielding gas between the IF and the DF, the DF slows until it reaches a velocity equal to that of the IF. If the DF initially propagates ahead of the IF-driven shock, the distance and column density between the IF and DF may in fact decrease in time (cf. §5.3; Hill & Hollenbach 1978).

### 5.1. Width of a Stationary Dissociation Front

In order to estimate the width of the DF, we first consider the structure of a stationary planar PDR that is illuminated by an FUV flux $F_{uv} = \chi F_H$ in the wavelength interval $912 - 1110$Å, where $F_{uv}$ is measured at the IF, which constitutes the surface of the PDR. We define $F_\lambda \equiv F_{uv}/(\Delta\lambda)_{uv}$ as the average flux per unit wavelength in the interval $(\Delta\lambda)_{uv} = 1110 - 912$Å$= 198$Å.

Let $N_2(x)$ be the column density of $H_2$ from the surface to some distance $x$ into the PDR, let $n_{\rm H} = 2n(H_2) + n(H)$ be the total hydrogen density, $\tau(x) = \sigma_{uv}N_{\rm H}(x)$ the dust optical depth to FUV photons within the PDR, and let us define $W = W_\lambda/\lambda = W_\nu/\nu$ as the dimensionless total equivalent width of the $H_2$ pumping lines.

The dissociation rate at a column density $N_2$ is

$$\zeta(N_2) = f_d \ \lambda \ F_\lambda \ \frac{dW}{dN_2} e^{-\tau} \ , \tag{21}$$

and a balance between $H_2$ formation and destruction in a dusty PDR requires

$$n(H_2) \ \zeta(N_2) \ = \ n(H) \ n_{\rm H} \ R_{gr} \ . \tag{22}$$

We have estimated the equivalent width for a gas with a moderate excitation temperature $T \sim 100$K and Doppler broadening of a few km/s, values typical in molecular clouds, and found that for $N_2 \lesssim 10^{14} \, \mathrm{cm}^{-2}$, $dW/dN_2 \simeq 5.14 \times 10^{-18} \, \mathrm{cm}^2$ and $f_d \simeq 0.14$. Although an individual line's equivalent width grows $\propto N_2^{1/2}$ for $N_2 \gtrsim 10^{16} \, \mathrm{cm}^{-2}$, the total dissociation rate approximately follows a power law for $10^{14} < N_2 < 10^{22} \, \mathrm{cm}^{-2}$ (Draine & Bertoldi, in preparation):

$$\zeta(N_2) \approx 4.6 \times 10^{-11} \, \chi \, e^{-\tau} \left( \frac{N_2}{10^{14} \, \mathrm{cm}^{-2}} \right)^{-3/4} \mathrm{s}^{-1} \, , \tag{23}$$

which in a simple form accounts for self-shielding, overlap (significant at $N_2 > 10^{20} \, \mathrm{cm}^{-2}$), and the variation of the dissociation fraction $f_d(N_2)$ because of the lines' saturation at different column densities. The steady-state solution for the molecular fraction $y \equiv 2n(\mathrm{H}_2)/n(\mathrm{H})$ is now

$$y = \frac{2R_{gr} n_{\mathrm{H}}}{\zeta(N_2)} = 1.30 \times 10^{-6} \left( \frac{n_{\mathrm{H}}}{\chi \mathrm{cm}^{-3}} \right) \left( \frac{R_{gr}}{3 \times 10^{-17} \mathrm{cm}^3 \mathrm{s}^{-1}} \right) \left( \frac{N_2}{10^{14} \, \mathrm{cm}^{-2}} \right)^{3/4} e^{\tau(x)} \, . \tag{24}$$

Let $L \equiv |d \ln y / dx|_{y=1}^{-1}$ characterize the width of the H/$\mathrm{H}_2$ transition in a stationary DF. From equation (24) and $|dN_2/dx|_{y=1} = n_{\mathrm{H}}/4$, the column density $n_{\mathrm{H}} L$ of the DF is then derived as

$$n_{\mathrm{H}} L = \sigma_{uv}^{-1} \left[ 1 + \phi^{-4/3} \right]^{-1} \, , \tag{25}$$

where

$$\phi = \phi_0 \, e^{-\tau_{pdr}} \, , \tag{26}$$

$$\phi_0 \equiv 12.4 \left( \frac{\chi}{n_{\mathrm{H}}/\mathrm{cm}^{-3}} \right) \left( \frac{3 \times 10^{-17} \mathrm{cm}^3 \mathrm{s}^{-1}}{R} \right) \left( \frac{\sigma_{uv}}{7.6 \times 10^{-22} \, \mathrm{cm}^2} \right)^{3/4} \, , \tag{27}$$

and $\tau_{pdr} = \tau(y = 1)$ is the UV opacity of the PDR up to the point where half the H is in molecular form. Numerical models of stationary PDRs (Draine & Bertoldi, in preparation) have

$$\tau_{pdr} \approx \ln(1 + 2.7\phi_0) \, . \tag{28}$$

Thus $\phi_0 \propto \chi/n_{\mathrm{H}}$ is a dimensionless measure for the width of the PDR, which derives from a balance of $\mathrm{H}_2$ formation and dissociation. Now the column density of the DF is

$$n_{\mathrm{H}} L \approx \sigma_{uv}^{-1} \left[ 1 + \left( 2.7 + \phi_0^{-1} \right)^{4/3} \right]^{-1} \, . \tag{29}$$

Comparison with numerical models (Draine & Bertoldi, in preparation) indicates that equation (29) tends to underestimate $n_{\mathrm{H}} L$, but is accurate to within a factor 2 for $\phi_0 > 0.01$. Note that this tendency to underestimate $n_{\mathrm{H}} L$ for stationary fronts is compensated for by the fact that propagating DFs are in fact narrower than stationary DFs with the same $\phi_0$.

## 5.2. Non-Equilibrium Condition

Now we are able to assess the importance of transient photochemistry on the surface of a photoevaporating cloud where both the IF and DF propagate into the cloud at the velocity $v_{IF}$ (eq.[17]). We now also take into account that the incident UV flux is attenuated in the ionized evaporation flow by $\tau_{uv} \simeq s\tau_{ly}$ before reaching the surface of the PDR. With equations (9), (15), and (16), and the identity $(S_{ly}/4\pi R^2 n_0 c_2)^2 = 2\psi\delta$ we obtain

$$\phi_0 = 6.62 \, v_{A0,5} \delta^{1/4} \sigma_{-21}^{-1/4} s^{3/4} \left(\frac{S_{uv}}{S_{ly}}\right) \frac{\tau_{ly} e^{-\tau_{ly}(s-3/4)}}{(1+\tau_{ly}/3)^{3/4}} \begin{cases} 1 & \text{for } R \gg r_c, \\ 0.85(r_c/R)^{1/4} & \text{for } R \ll r_c. \end{cases} \quad (30)$$

Advection will be important in the DF when $v_{DF}[dn(\text{H}_2)/dx] > R_{gr}n_{\text{H}}n(\text{H})$. Evaluated where $y = 1$, and using the estimate (29) for a stationary front, this becomes

$$v_{DF} > 4R_{gr}n_{\text{H}}L = 1.6 \left(\frac{7.6 \times 10^{-22}\,\text{cm}^2}{\sigma_{uv}}\right) \left[1 + (2.7 + \phi_0^{-1})^{4/3}\right]^{-1} \text{ km s}^{-1}. \quad (31)$$

With $v_{DF} = v_{IF} = S_{ly}/(4\pi R^2 qn_1)$ [cf. eq.(17)], and $q$ given by equation (12) and $n_1$ expressed through $\phi_0 \sim \chi/n_1$ (eq.27), we derive that advection will be important when

$$\phi_0 + \phi_0 \left(1.53 + \phi_0^{-1}\right)^{4/3} > 0.160 \left(\frac{s\sigma_{-21}}{0.76}\right)^{-1/4} \left(\frac{S_{uv}}{S_{ly}}\right) \left[1 + \frac{54.5\mathcal{F}}{\sigma_{-21}} \frac{\tau_{ly}}{1+\tau_{ly}/3}\right] e^{-\tau_{ly}(s-1)}. \quad (32)$$

The LHS attains a minimum of 3.06 when $\phi_0 = 0.106$, whereas for the typical dust properties we chose, the RHS attains a maximum value of $(10.6, 9.8)(S_{uv}/S_{ly})$ (corresponding to the limits $R/r_c \gg 1$ and $\ll 1$), when $\tau_{ly} = 2.26$. Thus the propagation of the IF and DF always has an important effect on the structure of the DF when the illuminating source has $S_{ly}/S_{uv} \gtrsim 3.5$. Since O stars tend to have flux ratios smaller than this value, whether or not propagation effects are important depends on the specific values of $\phi_0$ and $\tau_{ly}$. In Figures 3 we show for what range of values for $\phi_0$ and $\tau_{ly}$ condition (32) is fulfilled. Only inside the oval lines, which correspond to different values of $S_{ly}/S_{uv}$, is the DF slow enough that the chemical abundances reach near-equilibrium values – everywhere else transient photochemistry dominates the structure of the DF. In Figures 4 we display the same, but with $\phi_0$ replaced by the Strømgren parameter $\delta$. In Figures 5 we display the same parameter space, but only for the cases $S_{ly}/S_{uv} = 0.2$, corresponding to a late O star, and $S_{ly}/S_{uv} = 2$, corresponding to a very early or young O star, where in addition, we marked the regions of parameter values where the IF and DF overlap, where the cloud is instantly ionized by an R-type IF, and where initially the DF propagates ahead of the shock.

## 5.3. Trapping of the DF in the Shocked Layer

When an O star ignites in a clumpy GMC, the tenuous interclump gas is quickly ionized, leaving the dense molecular clumps exposed to the strong ionizing radiation that slowly evaporates and compresses them (Bertoldi 1989a,b; Bertoldi & McKee 1990). The clumps may be located deep inside the H II region for a long time – they are not necessarily located at the edge of an ionization-bounded H II region – so that the attenuation of UV photons due to the tenuous H II region gas between the star and the clump may not be significant. The PDRs on the surfaces of such clumps differ therefore from those discussed by Hill & Hollenbach (1978), who investigated the interplay of a DF and IF at the edge of an ionization–bounded H II region.

When a clump is first exposed to the strong dissociating and ionizing radiation, an IF rapidly ionizes a surface layer of width $2\delta r_c$ on the side of the cloud facing the star. Due to the high thermal pressure of this "Strømgren layer," it expands both into the H II region and the cloud, thereby driving a shock into the cloud that assembles a layer of compressed gas between the ionization and shock fronts (Bertoldi 1989b). After a time $\approx r_c/2c_2$, a steady evaporation flow is established and the shocked layer's column density near the apex of the cloud surface grows steadily. The DF may initially advance ahead of the shock. This occurs when the shock speed is lower than the speed with which the DF can propagate into the unshocked cloud, i.e. when $v_{DF}(n_0) > v_s$, which with equations (15) and (18) [substituting $n_0$ for $n_1$ and making use of eqs.(9), (10), and (12)] requires

$$\delta > 1.1 \times 10^{-4} \left(\frac{S_{ly}}{S_{uv}}\right)^4 \left(1 + \frac{\tau_{ly}}{3}\right)^3 \left(\frac{\sigma_{-21}}{\tau_{ly}}\right)^4 e^{4\tau_{ly}(s-3/4)} \times \begin{cases} 1 & \text{for } R \gg r_c, \\ 1.86 \ R/r_c & \text{for } R \ll r_c. \end{cases} \quad (33)$$

The broken lines in Figures 5 delineate the range in parameter space where condition (33) is met. Even if the DF initially propagates ahead of the SF, it may not reach its equilibrium extent because the continual growth of the compressed layer and the enhanced $H_2$ formation rate therein steadily decreases the equilibrium width of the PDR, until the shock engulfes the DF and confines it to the compressed layer between the IF and shock. While the DF propagates ahead of the shock, its speed relative to the unperturbed gas is about half the speed of the shock, and is therefore much higher than its speed after it is trapped in the shocked layer. The advective effects discussed in §5.2 are therefore even more pronounced throughout the somewhat intricate initial phase. Once the PDR is trapped in the shocked layer, we found above that over a large region of parameter space, transient photochemistry due to strong advection may dominate the structure of the PDR.

## 6. Examples

### 6.1. M17 Northern Bar

To quote an example of a photoevaporating and dissociating globule where we would expect transient photochemistry to be important, we consider the globular structure observed in detail by Chrysostomou et al. (1993) and Gatley & Kaifu (1987: Fig.7) in the northern bar of M17. Estimating a $25'' = 0.27$ pc radius of curvature and a projected distance of 1.9 pc from the group of one O4 and two O5 stars with total $S_{49} \simeq 20$ (Felli et al. 1984), we derive $\log \psi \simeq 4.6$ and $\log \delta \simeq -2.3 - 2 \log(n_0/10^4 \, \mathrm{cm^{-3}})$. Assuming the uncompressed density of the molecular globule $n_0 \approx 10^4 \, \mathrm{cm^{-3}}$ (the compressed gas appears to be at about $10^5 \, \mathrm{cm^{-3}}$), we find that the DF and IF may have separated (estimating $S_{ly}/S_{uv} \simeq 1.5 - 2$), although their parameters fall well into the domain where transient photochemistry is important.

An uncertainty in this estimate derives from our neglect of absorption of UV and Lyc photons in the H II region between the globule and the stars. Because of the strong clumping of the ionized gas and the complicated geometry of the H II region (Felli et al. 1984) it is difficult to resolve this issue. However, neglecting intervening attenuation, we would expect the peak density of the ionized gas streaming off the globule's IF to be $n_2 = S/(4\pi R^2 q v_2) \simeq 1700 \, \mathrm{cm^{-3}}$, where we computed $q \simeq 76$ and $\tau_{ly} \simeq 0.56$, assuming dense cloud dust properties. Our predicted value of the peak density agrees well with the $\sim 1000 \, \mathrm{cm^{-3}}$ estimate of Chrysostomou et al. (1992) from Br $\gamma$ observations, or with the density of $2400 \, \mathrm{cm^{-3}}$ determined by Felli et al. (1984) for clump #11 based on VLA radio continuum observations. This indicates that the ionizing flux arriving at the clump cannot be much lower than our estimate.

### 6.2. Gum Globules

The photoevaporating cometary globules in the Gum Nebula (cf. Bertoldi & McKee 1990) are believed to be photoionized by either $\zeta$ Pup=HD66811 (O4 I; Walborn 1972) or $\gamma^2$ Vel=HD68273 (WC8+O9I; van der Hucht et al. 1981). Because of their large distance ($\sim 50$ pc) from the ionizing star(s), the globules have small photoevaporation parameters, $\psi < 10^2$, and the outflows will not be very dusty, with $\tau_{ly} \lesssim 0.05$. Thus the condition (19) for merging of the IF and DF will be satisfied for $S_{ly}/S_{uv} \gtrsim 0.5$, which will certainly be satisfied for the ionizing star(s): these globules should therefore lack extended neutral H layers surrounding them. Note that even at $d \approx 50$pc $\zeta$Pup or $\gamma^2$Vel dominate the FUV starlight background.

### 6.3. Orion A: M42

A very prominent PDR is the one illuminated by the Orion Trapezium stars at the surface of the molecular cloud from which the Trapezium cluster emerged. Here the ionizing stars are very close to the cloud surface, so that $R \ll r_c$ and we need not worry about the actual cloud curvature, but only assume that the ionized gas is free to stream off the cloud surface. We assume our line-of-sight to be approximately normal to the cloud surface, which lies $R \approx 0.2$pc beyond $\theta^1$Ori C (Wen & O'Dell 1995). The ionizing flux from the Trapezium may actually have created a cup-like region (Pankonin, Walmsley & Harwit 1979; Wen & O'Dell 1995), so the planar boundary considered here is highly simplified.

The most luminous star in the Trapezium, $\theta^1$Ori C, is difficult to classify: Walborn (1981) reported variations in the apparent spectral type from O6 to O4 during 7 consecutive nights! To estimate the total ionizing flux from the Trapezium, we note that the 23GHz map of Wilson & Pauls (1984) has a peak brightness temperature of 24K, very nearly centered on $\theta^1$Ori C; the brightness temperature falls to about 25% of the peak at an angular distance of 2.1'. A uniform density ionized sphere of radius $R_s = 0.32$pc (2.2' at the assumed distance of 500pc), density $n_e = 2900$ cm$^{-3}$, and temperature $T = 9000$K, reproduces the peak surface brightness and produces 340 Jy at 23GHz, about 85% of the total measured flux of 400 Jy (Wilson & Pauls 1984). The recombination rate in this ionized sphere would be $9.5 \times 10^{48}$ s$^{-1}$, but the rate of emission of ionizing photons by the star must be substantially larger, since a large fraction of the ionizing photons will be absorbed by dust. The extinction curve toward $\theta^1$Ori C (Lee 1968) is characterized by $R_V = 5.5$ (Clayton, Cardelli & Mathis 1989), from which the extinction cross section at 912Å would be estimated to be $9.65 \times 10^{-22}$ cm$^2$; taking the absorption cross section $\sigma_{ly} \approx 5 \times 10^{-22}$ cm$^2$, and $n_e/n_H = 1.08$, the dust optical depth from center to edge is 1.33, so that absorption by dust is important. Following Petrosian, Silk & Field (1972), we estimate that ~64% of the ionizing photons are absorbed by dust; thus we would estimate $S_{ly} = 2.6 \times 10^{49}$ s$^{-1}$. The model is obviously imperfect: M42 is not spherically symmetric, there must be a stellar-wind-supported central cavity, there is additional ionization beyond 2.2', and the dust properties are uncertain. In the analysis below we take $S_{49} \approx 3$ for the Trapezium stars, mainly due to $\theta^1$Ori C.

If $\theta^1$Ori C radiates like a $T = 45000$ K blackbody with $S_{49} = 2.6$, then its luminosity $L = 4.2 \times 10^5 L_\odot$, consistent with spectral type O5.5 ZAMS (Panagia 1973). The other Trapezium stars ($\theta^1$ Ori A=B0.5V, $\theta^1$ Ori D=O9.5V; Carruthers 1969) raise the total luminosity to $L \approx 5 \times 10^5 L_\odot$. We assume the Trapezium to lie a distance $R \approx 0.2$pc in front of the molecular cloud; if a fraction $f_{IR}$ of the radiation from the star is converted to infrared emission, then the resulting integrated surface brightness of the cloud surface behind the Trapezium would then be $\sim L f_{IR}/16\pi^2 R^2 = 32 f_{IR}\,\mathrm{erg\,cm^{-2}\,s^{-1}\,sr^{-1}}$, in reasonable agreement with the measured $\sim 10\,\mathrm{erg\,cm^{-2}\,s^{-1}\,sr^{-1}}$ (Werner et al. 1976; Drapatz et al. 1983). Note that if $R = 0.1\mathrm{pc}$, as favored by Tielens & Hollenbach 1985, $L/16\pi^2 R^2$ would exceed the measured value by an order of magnitude.

From equation (1) we find $\psi R/r_c \simeq 6.7 \times 10^5$. With $\sigma_{ly} = 5 \times 10^{-22}\,\mathrm{cm^2}$, equation (11) becomes

$$\psi = 1.19 \times 10^5 \; \frac{\tau_{ly}^2 e^{\tau_{ly}}}{1 + \tau_{ly}/3} \times 0.61(r_c/R) \quad \text{for } R \ll r_c \; , \tag{34}$$

from which we obtain $\tau_{ly} \approx 1.65$, $q \approx 560$, and $n_2 v_2 \approx 9.7 \times 10^9\,\mathrm{cm^{-2}\,s^{-1}}$. Taking $v_2 = 1.6c_2$, we estimate the electron density at the ionization front to be $n_2 = 5300\,\mathrm{cm^{-3}}$. The density in the PDR has been estimated to be $n_1 = 2 \times 10^5\,\mathrm{cm^{-3}}$ (Tielens & Hollenbach 1985), and we assume an isothermal sound speed $c_1 = 2.5\,\mathrm{km\,s^{-1}}$, corresponding to $T_1 \approx 1000\,\mathrm{K}$. Rather than use the approximations (15) and (16), which are accurate only for a strong shock, we apply the momentum conservation jump condition across the ionization front:

$$\rho_2 \left[v_2^2 + c_2^2 + (1/2)v_{A2}^2\right] = \rho_1 \left[v_1^2 + c_1^2 + (1/2)v_{A1}^2\right] \; , \tag{35}$$

where a transverse magnetic field is assumed. Since $v_1 = (\rho_2/\rho_1)v_2$, and $v_{A2}^2 = (\rho_2/\rho_1)v_{A1}$, we may solve for $v_{A1}$:

$$v_{A1}^2 = 2 \; \frac{(\rho_2/\rho_1)(v_2^2 + c_2^2) - (\rho_2/\rho_1)^2 v_2^2 - c_1^2}{1 - \rho_2/\rho_1} \; . \tag{36}$$

The density ratio $\rho_1/\rho_2 = 38$ then implies $v_{A1} = 3.4\,\mathrm{km\,s^{-1}}$. Similarly, we may apply the momentum jump conditions to the shock front to obtain

$$v_s^2 = \frac{(\rho_1/\rho_0)c_1^2 - c_0^2 + (\rho_1/\rho_0 - \rho_0/\rho_1)v_{A1}^2/2}{1 - \rho_0/\rho_1} \; . \tag{37}$$

If we assume a shock compression ratio $\rho_1/\rho_0 = 2$ (i.e., preshock density $n_0 = 1 \times 10^5\,\mathrm{cm^{-3}}$) and $c_0 = 0.5\,\mathrm{km\,s^{-1}}$, we obtain $v_s = 6.5\,\mathrm{km\,s^{-1}}$. The preshock Alfvén speed $v_{A0} = v_{A1}(\rho_0/\rho_1)^{1/2} \approx 2.4\,\mathrm{km\,s^{-1}}$ is within the range 1–3 $\mathrm{km\,s^{-1}}$ believed to prevail in molecular clouds. The PDR gas would then be moving at a velocity $(1 - \rho_0/\rho_1)v_s = 3.2\,\mathrm{km\,s^{-1}}$ relative to the preshock molecular cloud. This is in agreement with the observed velocity difference of 3 $\mathrm{km\,s^{-1}}$ between the C109$\alpha$ line at $v_{LSR} = 11\,\mathrm{km\,s^{-1}}$ and the molecular gas at $v_{LSR} \approx 8\,\mathrm{km\,s^{-1}}$ (Jaffe & Pankonin 1978; Pankonin, Walmsley & Harwit 1979).

Will the IF and DF be separate or merged? The initial speed ratio is

$$\frac{v_{IF}}{v_{DF}} = \left(\frac{S_{ly}}{S_{uv}}\right) \frac{e^{s\tau_{ly}}}{2\epsilon\, f_d\, q} \; . \tag{38}$$

With $q = 560$, $\tau_{ly} = 1.65$, $s \approx 1.3$, and $f_d = 0.15$, this becomes $v_{IF}/v_{DF} = 0.10(S_{ly}/S_{uv})(0.5/\epsilon)$. If we assume that $\theta^1$Ori C radiates like a $T \approx 45000$ K blackbody, then $S_{ly}/S_{uv} \approx 2.4$. This ratio is higher than that implied by equation (20), but can be justified by the fact that very young stars tend to have higher effective temperatures and thereby higher Lyc to FUV flux ratios than averaged over their lifetime. Accounting for a population of later type O and B stars close to the cloud, we overall estimate $S_{ly}/S_{uv} \approx 2$. Therefore we estimate that $v_{IF}/v_{DF} \approx 0.2$, a value which is insensitive to the chosen value of $R$, but very sensitive to the ratio between the effective Lyc and FUV attenuation cross sections, $s$. With our chosen values, we find that the DF and IF will be able to separate until dust absorption in the PDR slows the DF to the speed of the IF. We also note that for an assumed distance $R = 0.2$pc from the star, the UV radiation field will be characterized by an intensity $\chi = (S_{uv}e^{-s\tau_{ly}}/4\pi R^2 F_H) = 2.6 \times 10^4$ – somewhat weaker than the UV intensity $\chi = 10^5$ postulated in the PDR model of Tielens & Hollenbach (1985).

With equation (27) we estimate $\phi_0 = 1.4$. From equation (31) advection will be important for $v_{DF} > 0.30$ km s$^{-1}$. The velocity of the IF – and therefore of the DF – is $v_1 = 0.48$ km s$^{-1}$ (eq.[17] with $n_1 = 2 \times 10^5$ cm$^{-3}$), implying that the DF is propagating rapidly enough that its structure is somewhat affected by transient photochemical effects. Note that the value we adopted for $n_1$ was based on modelling (Tielens & Hollenbach 1985) using a radiation field $\sim 4$ times stronger than our estimate. If $n_1$ is in fact smaller, $v_1$ would be larger, and the effects of advection on the abundances in the PDR enhanced. The strong absorption of Lyc radiation in the H II region and evaporation flow makes the average energy of the Lyc photons arriving at the cloud surface very high, which would probably justify a lower value of $\sigma_{ly}$ than the one we adopted.

## 7. Discussion and Summary

We have shown that existing models of stationary PDRs are often inapplicable to dense photodissociation regions around molecular clouds that are directly exposed to the ionizing and dissociating radiation from massive, early-type stars. Stars of early spectral types have sufficiently large values of $S_{ly}/S_{uv}$ that, for a wide range of parameters, the DF is not able to separate from the IF. When this occurs, the IF and DF form a merged structure, and the UV pumping and photodissociation of $H_2$ occurs to a considerable extent in gas which is also undergoing photoionization. Consequently, most dissociating FUV radiation as well as some of the Lyc radiation is absorbed by $H_2$, not dust, and the efficiency for conversion of starlight into $H_2$ line emission should be much enhanced over that of static, thick PDRs.

Another consequence is that the UV pumping of $H_2$ will take place in a region where intense Lyman-$\alpha$ radiation is present. If the Lyman-$\alpha$ is sufficiently intense, it may affect the IR spectrum of vibrationally-excited $H_2$, since the Lyman-$\alpha$ can selectively depopulate the $v=2, J=5$ and $v=2, J=6$ levels (Shull 1978; Black & van Dishoeck 1987). Whether the Lyman-$\alpha$ intensities will be large enough to significantly affect the IR emission spectrum is unclear, but merits investigation. This may prove to be the explanation for the reported weakness of the 2–1S(3) line in the spectrum of NGC 6240 (Lester, Harvey & Carr 1988).

$H_2^+$ will be present in the merged IF/DF, and it is possible that detectable levels of vibrational emission may be produced. The $H_2^+$ will react with $H_2$ to produce $H_3^+$, which also may emit detectable levels of vibration-rotation emission. Theoretical modelling of merged ionization/dissociation fronts is currently underway (Bertoldi & Draine, in preparation).

Even when the DF is able to propagate ahead of the IF, we find that the DF will often propagate too fast for stationary PDR models to provide a good approximation to the chemical and thermal structure of the PDR. When this occurs, we expect that the temperature of the molecular gas will be increased due to the increased rate of UV pumping followed by dissociation (about $\simeq 15\%$ of the time) or collisional deexcitation (for densities $n \gtrsim 10^4 \, \mathrm{cm}^{-3}$). If the density is high enough for collisional deexcitation to be effective, the DF may produce intense $H_2$ vibrational line emission with a near-thermal spectrum. Stationary PDR models appear able to convert no more that $7 \times 10^{-5}$ of the energy output from a star into emission in the $H_2$ 1–0S(1) line, in which case PDRs are unable to account for the intense emission observed from NGC 6240 (cf. Draine & Woods 1990). It appears possible that propagating DFs may attain much higher efficiencies for conversion of starlight into $H_2$ line emission.

We predict that in several famous dissociation regions such as M17N and Orion, non-equilibrium conditions persist in the dissociation fronts and their PDRs may not be well developed, implying smaller H I column densities than predicted by equilibrium models. Models of propagating ionization-dissociation fronts are required to theoretically describe these regions. Such models are currently under development and will be described in future papers.

We thank D. Hollenbach for valuable discussions and criticism, and the referee, A. Sternberg, for helpful comments. This work was in part supported by the Deutsche Forschungsgemeinschaft and by NSF grants AST–9017082 and AST–9319283.

Figure 1

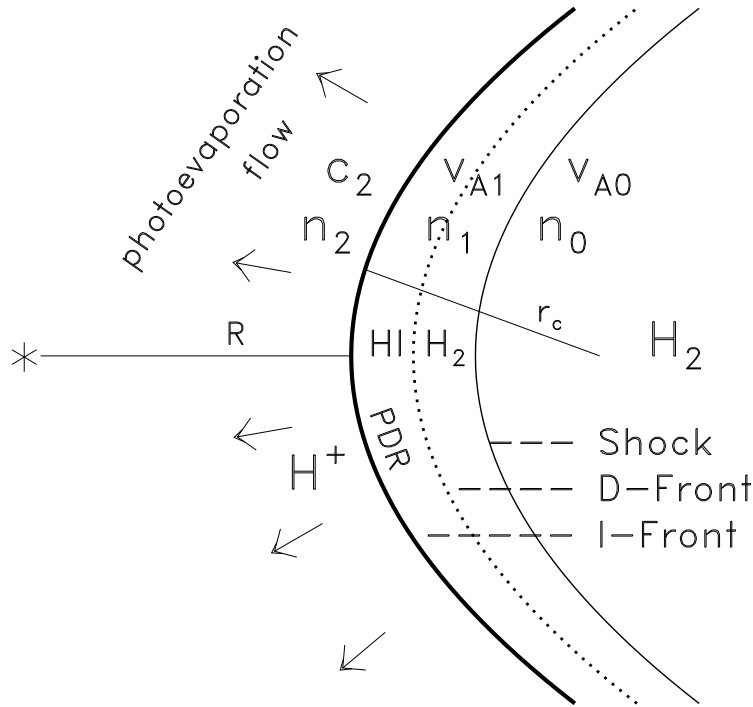

Fig. 1.— Schematic rendering of conditions on the surface of a molecular cloud exposed to ionizing and dissociating radiation from a single star: the compressed gas layer between the ionization and shock fronts propagates into the cloud, driven by the pressure of the ionized gas that streams off the surface with an effective velocity $v_2 \simeq 1.6 c_2$, surface density $n_2$ and sound speed $c_2$. The dissociation front here is trapped in the compressed layer, but may initially propagate ahead of the shock.

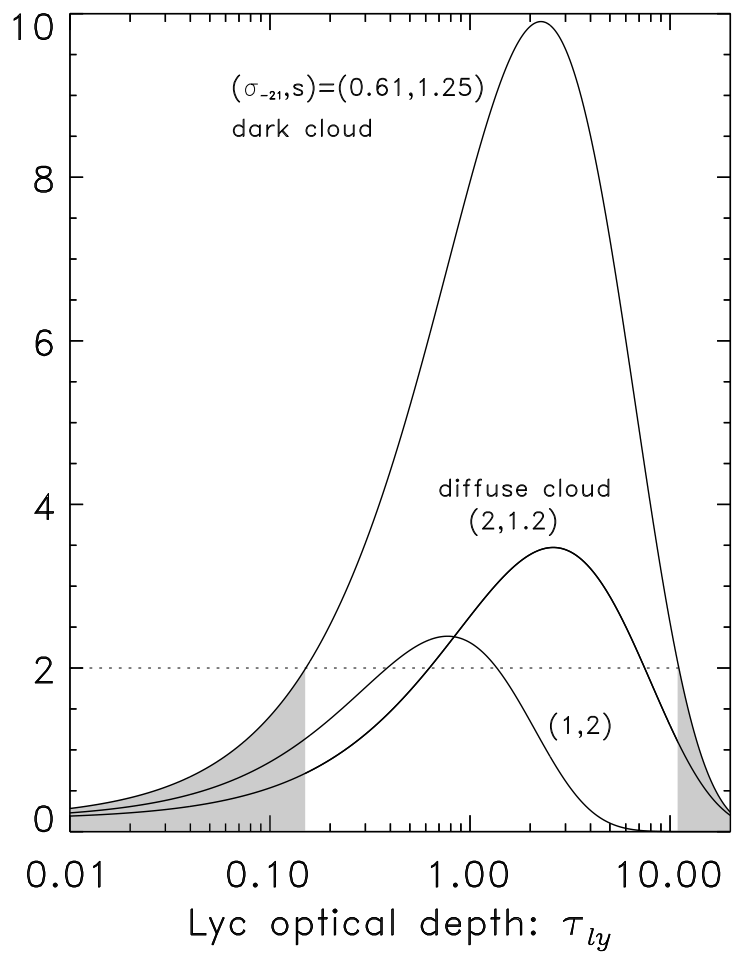

Fig. 2.— Evaluating the condition under which the ionization and dissociation fronts merge: the right-hand-side of eq.(19) as a function of the Lyc optical depth of the photoevaporation flow, $\tau_{ly}$, for different combinations of $\sigma_{ly} \equiv \sigma_{-21} 10^{-21} \mathrm{cm}^2$ and $s \equiv \sigma_{uv}/\sigma_{ly}$. For, e.g., $S_{ly}/S_{uv} = 2$ and dark cloud dust properties, the IF and DF merge for $\tau_{ly} < 0.15$ (cf. vertical line in Fig.5b) or $\tau_{ly} > 11$.

Figure 3a

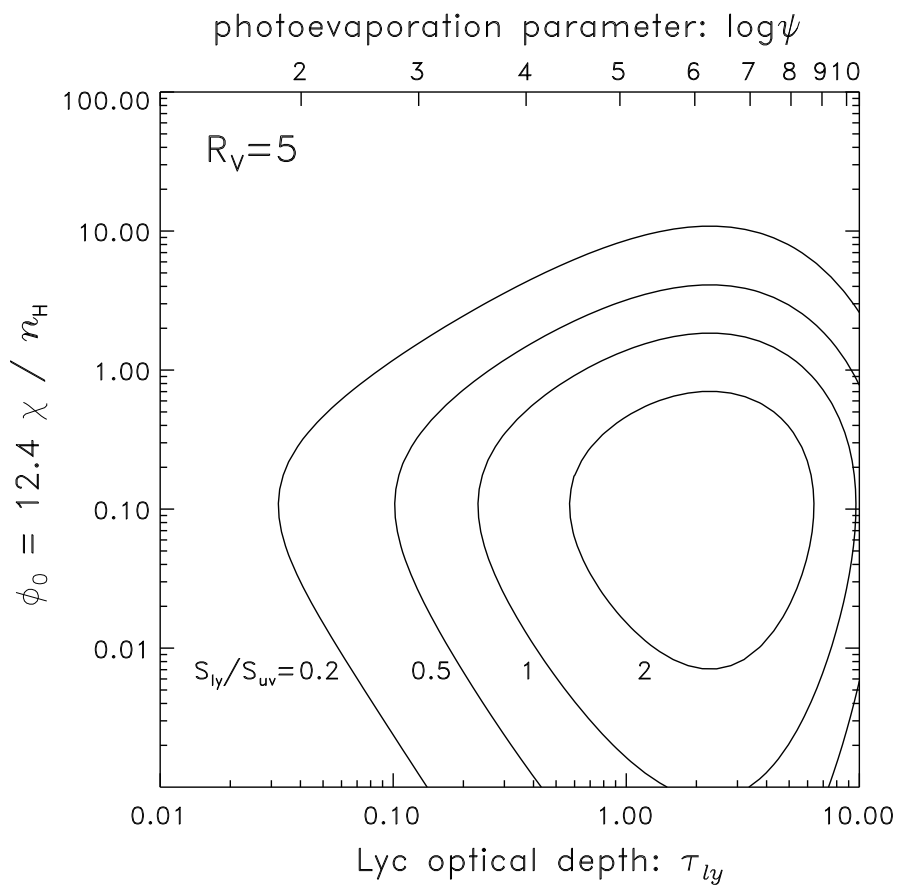

Fig. 3.— Cloud parameter space for $R \gg r_c$: $\phi_0$ is a measure for the width of the PDR and strength of the UV radiation field, the photoevaporation parameter, $\psi$, is a measure for the Lyc opacity, $\tau_{ly}$, of the evaporation flow. Equilibrium abundances are established in the PDR only inside the circular lines, which each corresponds to a different ratio of the Lyc to FUV flux incident on the cloud. (a) Dark cloud dust properties, $\sigma_{-21} = 0.61$, $s = 1.25$, (b) diffuse cloud dust, $\sigma_{-21} = 2$, $s = 1.20$.

Figure 3b

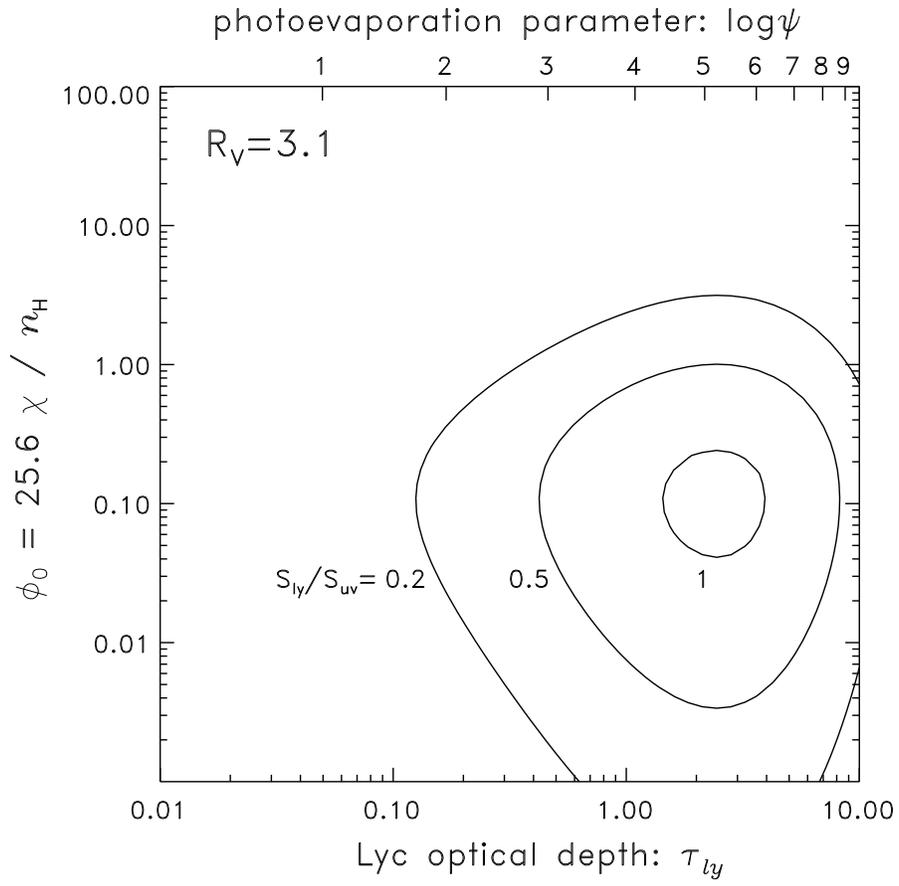

Figure 4a

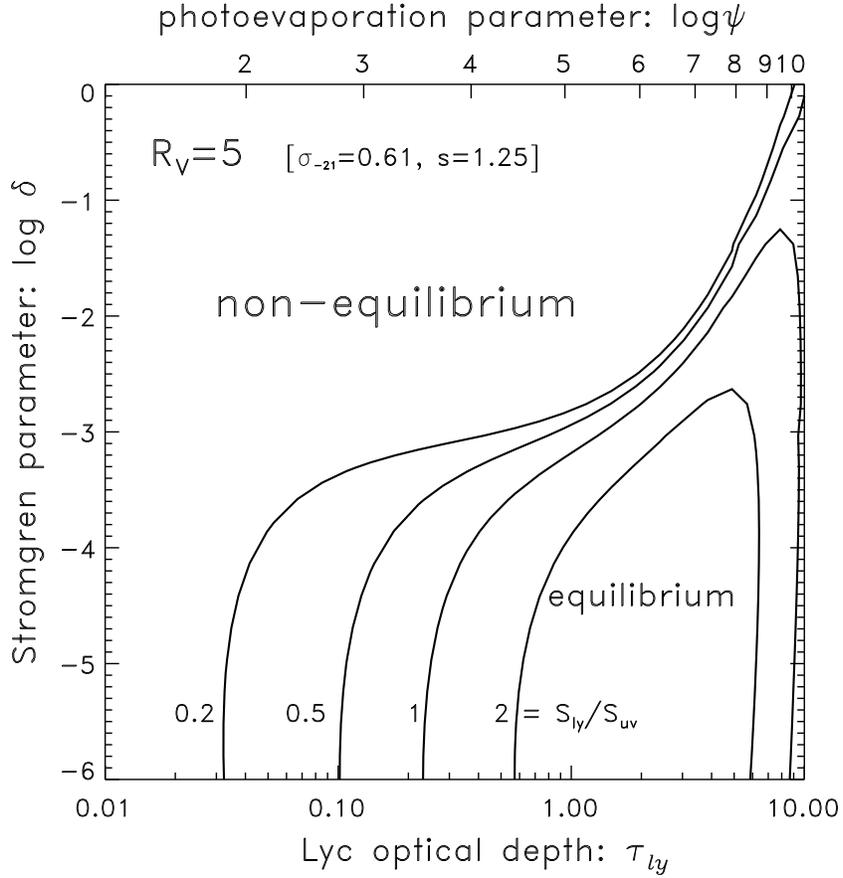

Fig. 4.— Cloud parameter space for $R \gg r_c$, $v_{A0} = 2\,\mathrm{km\,s^{-1}}$, and (a) dark cloud dust properties, $R_V = 5$, (b) diffuse cloud dust properties, $R_V = 3.1$. The regime where the DF propagates too rapidly for equilibrium PDR models to apply (non-equilibrium) is separated from the regime where equilibrium abundances are established in the PDR by lines corresponding to different ratios of the Lyc to FUV flux incident on the cloud.

Figure 4b

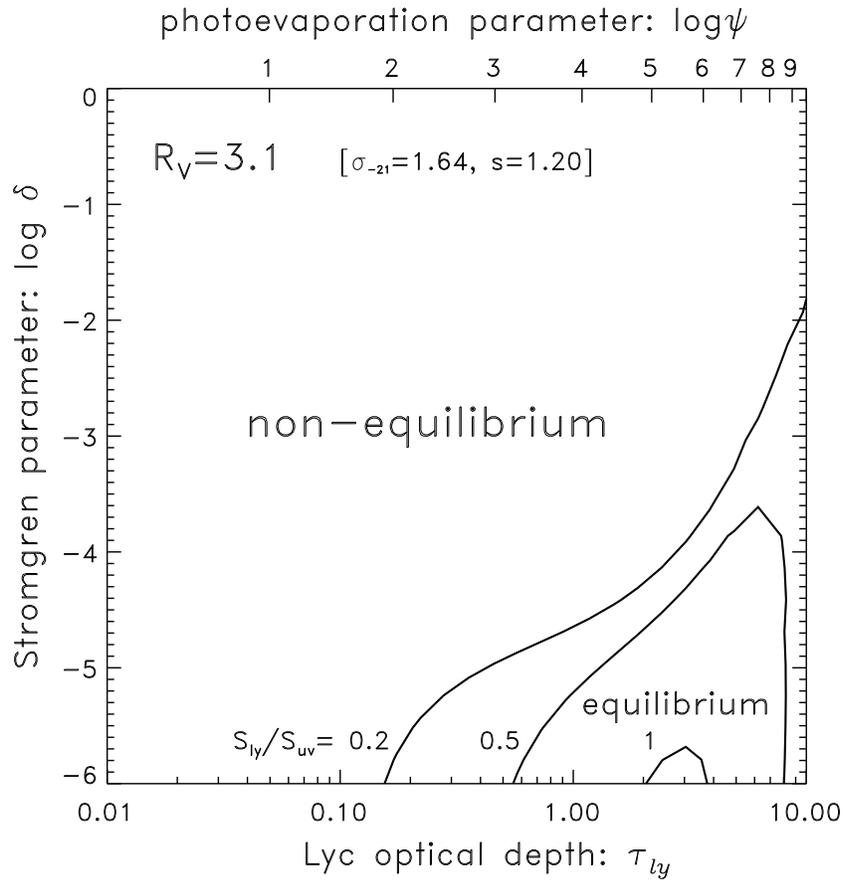

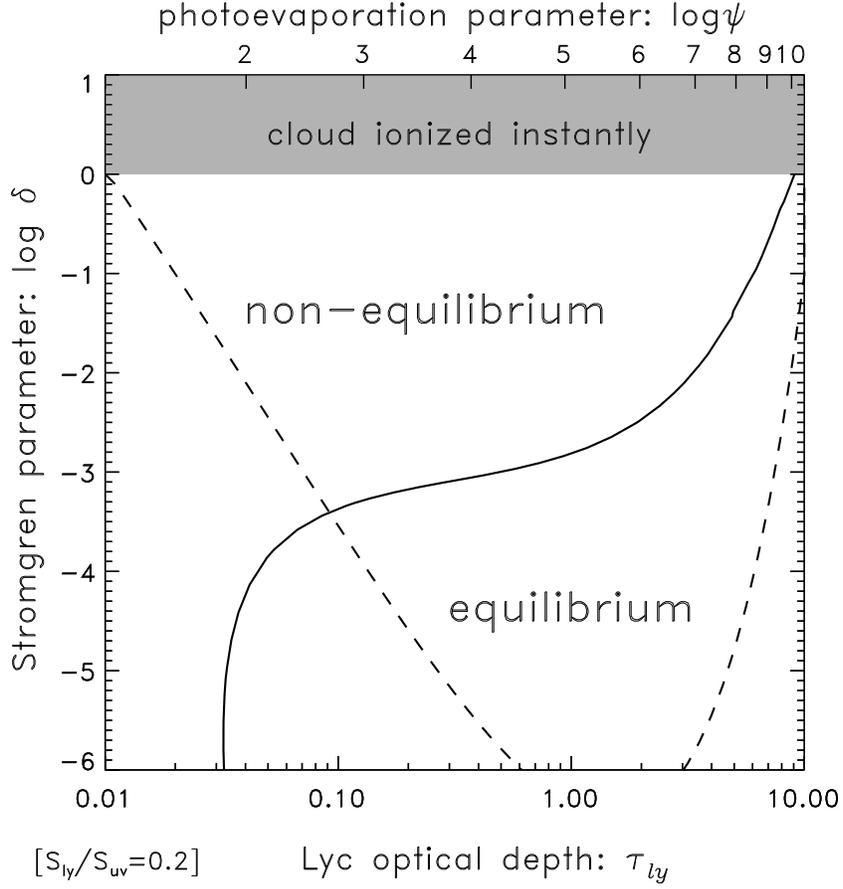

Fig. 5.— Cloud parameter space for $R \gg r_c$ for dark cloud dust, $\sigma_{-21} = 0.61$, $s = 1.25$, and $v_{A0,5} = 2$: the Strømgren number, $\delta$, is a measure of the dynamical impact of the photoevaporation on the cloud: clouds with $\delta > 1$ are rapidly ionized before a shock can form, shocks are driven into the cloud for smaller values of $\delta$, and for $\delta \lesssim 10^{-5}$ the cloud is dynamically unaffected by the photoevaporation. (a) $S_{ly}/S_{uv} = 0.2$, corresponding to a $\approx 18 M_\odot$ star, (b) $S_{ly}/S_{uv} = 2$, corresponding to a $\approx 60 M_\odot$ star. Only for cloud parameters within the "equilibrium" area are equilibrium abundances established in the PDRs, elsewhere transient photochemistry dominates. Clouds to the left of the vertical line in Fig.5b have merged IF/DF structures [eq.(19)] and a PDR does not form. The region above the broken line is where the DF initially moves ahead of the shock front. For $R \ll r_c$, one obtains almost identical diagrams if $\delta$ is replaced by $\delta r_c/R$ and $\psi$ by $\psi R/r_c$.

Figure 5b

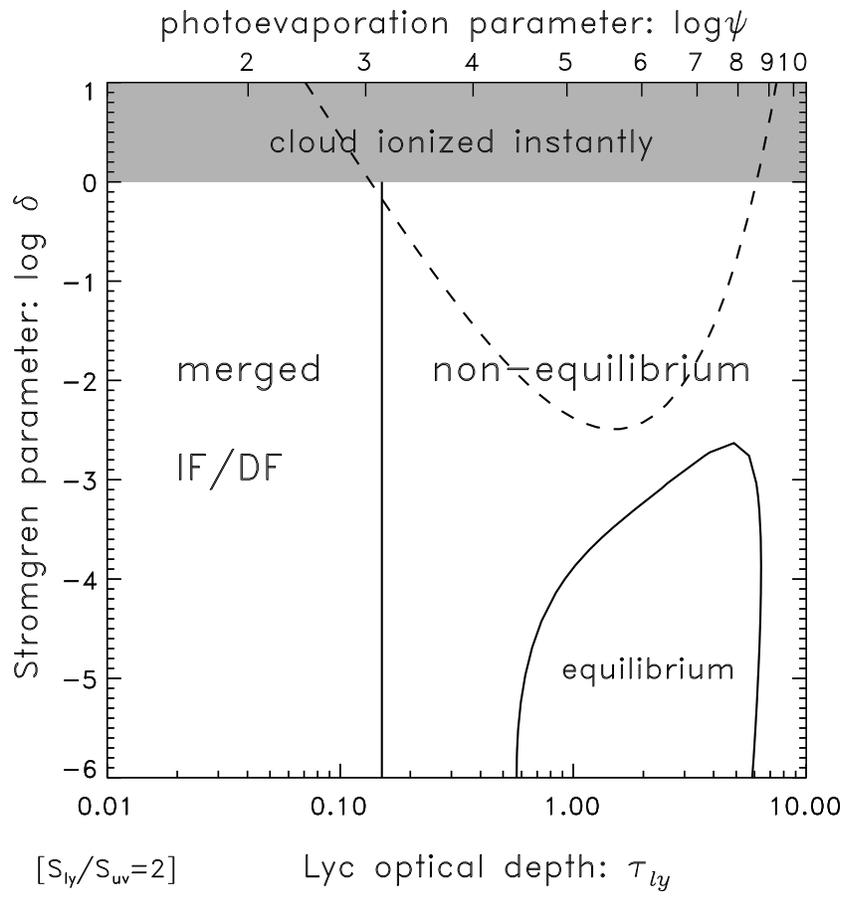